# Network Structural Dependency in the Human Connectome Across the Life-Span

Markus D. Schirmer[a,b,c,1*], Ai Wern Chung[d*], P. Ellen Grant[d] and Natalia S. Rost[a]

[a]Stroke Division & Massachusetts General Hospital, J. Philip Kistler Stroke Research Center, Harvard Medical School, Boston, USA
[b]Computer Science and Artificial Intelligence Lab, Massachusetts Institute of Technology, Cambridge, USA
[c]Department of Population Health Sciences, German Centre for Neurodegenerative Diseases (DZNE), Germany
[d]Fetal-Neonatal Neuroimaging & Developmental Science Center, Boston Children's Hospital, Harvard Medical School, Boston, MA, USA

*Authors contributed equally.

## Abstract

Principles of network topology have been widely studied in the human connectome. Of particular interest is the modularity of the human brain, where the connectome is divided into subnetworks and subsequently changes with development, aging or disease are investigated. We present a weighted network measure, the Network Dependency Index (NDI), to identify an individual region's importance to the global functioning of the network. Importantly, we utilize NDI to differentiate four subnetworks (Tiers) in the human connectome following Gaussian Mixture Model fitting. We analyze the topological aspects of each subnetwork with respect to age and compare it to rich-club based subnetworks (rich-club, feeder and seeder). Our results first demonstrate the efficacy of NDI to identify more consistent, central nodes of the connectome across age-groups, when compared to the rich-club framework. Stratifying the connectome by NDI led to consistent subnetworks across the life-span revealing distinct patterns associated with age where, e.g., the key relay nuclei and cortical regions are contained in a subnetwork with highest NDI. The divisions of the human connectome derived from our data-driven NDI framework have the potential to reveal topological alterations described by network measures through the life-span.

## Introduction

Network theoretical principles have been readily applied to the human connectome to investigate its structural and functional organization. Typically, a brain network comprises of nodes representing brain regions that are connected by edges representing either reconstructed white matter pathways from diffusion-weighted magnetic resonance imaging (dMRI) or functional correlations from functional MRI (fMRI). Furthermore, edges between two nodes may be weighted to reflect the strength of the connection (Bullmore and Sporns, 2009; Fornito et al., 2013). There has been an ongoing quest to identify the regions in brain networks that are critical for efficient network functioning (see e.g. (Hagmann et al., 2008)), as well as defining modules, or sets of nodes, that distinguish themselves from other nodes in the connectome. One network organizational principle that has been established in the human connectome is the rich club (RC) - a core subnetwork of brain regions that are strongly connected to form a high-cost, high-capacity backbone. The RC has been shown to be critical for effective communication in the connectome (van den Heuvel et al., 2012; Van Den Heuvel and Sporns, 2011) and has been studied in healthy subjects (Grayson et al., 2014; Schirmer and Chung, 2018; Van Den Heuvel and Sporns, 2011; Zhao et al., 2015), with development (Ball et al., 2014) and in disease (Collin et al., 2013; Daianu et al., 2015; Ray et al., 2014). While an important topological aspect in brain networks, defining RC membership is not straight-forward. Typically, an RC-regime is identified over a range of degrees where the RC coefficient is significantly greater than a distribution of equivalent RC coefficients computed from random networks. From this range, a single degree, k, is chosen and nodes with a degree greater or equal to k are deemed to form the RC subnetwork. The lack of consensus in choosing this k-th threshold is in part due to the variations in network construction and weightings employed in the field, leading to great variability in RC coefficients (Van Den Heuvel and Sporns, 2011). As such, the neuroscience community has yet to establish an optimal way of defining RC nodes.

In this work, we present an alternative for defining regions of the brain that are integral for network efficiency, and which can be consistently identified across the life-span, based on the network dependency index (NDI; (Woldeyohannes and Jiang, 2018)). First introduced in network communication science for critical node detection, NDI quantifies a node's importance as defined by the impact on the network's performance given the node's failure (or removal) from the system - simply put, NDI measures the dependency of the network on any given node. It also has the advantage of evaluating nodal importance by incorporating the network's efficiency for information transport as measured by topological distance. Ordering brain regions by NDI then allows us to stratify the connectome into subnetworks according to nodal importance. We present a comparative investigation between RC and NDI frameworks for detecting a core subnetwork and their corresponding 'peripheral' subnetworks. We demonstrate the efficacy of NDI to identify more consistent, central nodes of importance in the human connectome by applying our framework on a large, open-source, normal population from the NKI-Rockland study with an age range of 4 to 85 years. By doing so, we derive reference values evaluating NDI with age from which to contextualize our findings with other literature in the field. In the following sections, we describe both RC and our NDI framework to define a core subnetwork from which to further stratify the entire connectome. In Results, we compare the core regions detected by both frameworks, and their corresponding network characteristics in relation to age. We end with a Discussion of our main NDI findings in comparison to the RC formalism.

# Materials and Methods

## Study design and patient population

In this work we utilize data from the NKI-Rockland life-span study (Nooner et al., 2012). Preprocessed connectome data were obtained from the USC Multimodal Connectivity database[1]. MRI acquisition details are available elsewhere (Brown et al., 2012). In brief, a total of 196 connectomes of healthy participants are computed, from 3T dMRI acquisitions (64 gradient directions; TR=10000ms; TE=91ms; voxel size=2mm$^3$; *b*-value=1000 s/mm$^2$). Following eddy current and motion correction, diffusion tensors are modelled and deterministic tractography performed using fiber assignment by continuous tracking (angular threshold 45°; (Mori et al., 1999)). Regions of interest (ROI) are based on the Craddock atlas (Craddock et al., 2012), resulting in 188 ROIs and connections are weighted by the number of streamlines connecting pairs of ROIs. Here, we normalize each connectome by the maximum streamline count for each subject so that the connection weights $w_{ij}$ within each subject are $w_{ij} \in [0,1]$.

---

[1] http://umcd.humanconnectomeproject.org

For part of our analysis, we divide the 196 participants into four age groups: $U20 \leq 20$ years, 20 years $< U40 \leq 40$ years, 40 years $< U60 \leq 60$ years, 60 years $< O60 \leq 80$ years. Four subjects were above 80 years old (81, 82, 83 and 85 years). As there were only four subjects, we included them in the O60 group. Table 1 characterizes the study cohort and groups.

*Table 1: NKI-Rockland life-span study cohort characterization and their stratification by age (in years).*

|  | **Overall** | **U20** | **U40** | **U60** | **O60** |
|---|---|---|---|---|---|
| **N** | 196 | 53 | 67 | 47 | 29 |
| **Age (Mean (SD))** | 35.0 (20.0) | 13.8 (4.1) | 27.4 (5.9) | 47.4 (5.4) | 71.0 (6.8) |
| **Sex (Male; %)** | 58.1 | 54.7 | 56.7 | 72.3 | 44.8 |

The available connectivity matrices include both cortical and subcortical structures. Before analysis, regions in the brainstem and cerebellum were removed from the network, resulting in 170 nodes covering 46 bilateral, anatomical regions (see Table S1 in Supplementary Materials for distribution of nodes by anatomical region).

## Group and cohort connectomes

A group-averaged connectome of weighted matrices can be computed in two steps (Van Den Heuvel and Sporns, 2011). First, we calculate a binarized, group-average adjacency matrix by retaining edges that are present in at least 90% of the subjects in each group, thus preserving connections which can be reliably identified across the group/cohort. Weights are subsequently added to the group-averaged adjacency matrix by taking the average weight of each connection across the group, generating a weighted group-averaged connectome $W_{group}$. In addition to calculating a connectome for each age group, we also compute a cohort-based connectome, $W_{cohort}$, across all 196 subjects.

## Rich-club framework

We utilize $W_{group}$ to subsequently calculate the weighted RC parameter $\Phi_{group}(k)$ (Opsahl et al., 2008), where *k* denotes the degree of nodes. The RC parameter $\Phi_{group}(k)$ is normalized relative to a set of comparable random networks of equal size and with similar connectivity distribution. Here, we generate 1000 random networks while preserving weight, degree and strength distributions of $W_{group}$ (Rubinov and Sporns, 2010). For each of these random realizations of the graph, we calculate the weighted RC parameter $\Phi_{rand}(k)$. Finally, the normalized weighted RC parameter is calculated as

$$\Phi_{group}^{norm}(k) = \frac{\Phi_{norm}(k)}{\Phi_{rand}(k)}$$

For this metric, $\Phi_{group}^{norm}(k) > 1$ denotes the presence of a rich-club. In our analysis, we select

$$k_{max}^{group}: max(k), \text{ for which } \Phi_{group}^{norm}(k) > 1,$$

as the degree of the RC nodes of a given group, which allows us to determine the RC members with a degree of at least *k*.

Rich-club subnetworks

For each group connectome, we first determine $k_{max}^{group}$, as implemented in the Brain Connectivity Toolbox (Rubinov and Sporns, 2010). Statistical significance of $\Phi_{group}^{norm}(k) > 1$ is assessed by performing a right-sided t-test for each k (p<0.05 after Bonferroni correction, where the number of tests equals the maximum degree within each group connectome). After establishing the RC nodes, remaining nodes in the group connectome were categorized according to their relationship to the RC subnetwork. All non-RC nodes were identified as Feeders (*F*) if they were connected to an RC node, or Seeders (*S*) if they shared no connection with an RC node (Schirmer and Chung, 2018).

Network dependency index framework

The NDI score has recently been described in unweighted networks (Woldeyohannes and Jiang, 2018). Here, we extend their formalism for use in weighted networks. Given a connectivity matrix $W_{group}=\{w_{ij}\}$ in a network G with n nodes, we first calculate the topological distance matrix D between all node pairs using the inverse of the connection strength $w_{ij}$ between nodes i and j as an initial topological distance. The information measure $(I^n)_{ij}$ between nodes i and j can then be calculated as $1/D_{ij}$, resulting in an information measure matrix $I^n$. As a next step, we normalize $I^n$ by the maximum information measure $I_{max} = max(w_{ij})$. $I_{max}$ can be defined for each connectivity matrix W individually, however, as we compare NDIs across groups, we define it as the maximum of $I_{max}$ across all 196 connectivity matrices. In case of disconnected components, the maximum loss of information $I_{max}$ thereby becomes comparable across connectomes.

To determine the NDI of node m, we first create a subnetwork $G^{-m}$, by removing node m. We subsequently calculate $(I_{-m}^n)_{ij}$ for all remaining node pairs (*i,j*) in $G^{-m}$. If removal of node m results in disconnected components, some path-lengths may become infinite. For such paths, the loss of information measure is set to 1. Finally, each node $i \in G^{-m}$ is assigned an accumulated loss in information measure

$$\Delta I_i = \sum_j (I^n)_{ij} - (I^n_{-m})_{ij}.$$

The NDI of node m is then given as the mean of $\Delta I_i$ over all nodes i in $G^{-m}$, where the closer the value is to 1, the higher the information loss and subsequently the more important node m is to the network. This analysis is then repeated for all nodes in the network, resulting in an n×1 dimensional feature vector of NDI scores for the network.

### NDI subnetworks

For each group connectome, we calculate its NDI scores. We determine regions of importance by ranking their NDI score from highest to lowest for comparison with RC nodes. Additionally, we model the natural-log-transformed NDI of each group connectome, using a Gaussian Mixture Model with g Gaussian distributions ($GMM_g$). Using the halfway point between the Gaussian centers, we define g+1 subnetworks (with one additional subnetwork for nodes with NDI=0), referred to as *Tiers*, of decreasing NDI importance (from Tier 1, …, g+1, where Tier 1 contains nodes with greatest NDI scores).

## Statistical analysis

### Network organization by age group

Our first analysis is to investigate the consistency and distributions of NDI scores for each Tier in relation to rich-club subnetworks. Consistency of NDI across age groups is subsequently tested using Spearman's rank correlation coefficient. Additionally, to further assess this consistency, we generate confusion matrices, where nodal assignments to subnetworks/Tiers are compared between age groups. To characterize the topology of the subnetworks derived from each framework, we calculate three network measures (Rubinov and Sporns, 2010), namely transitivity (T), global efficiency (E) and assortativity (a).

### Age-dependent subnetwork trends within the cohort

We also investigate the associations of each network measure with age in both RC- and NDI-defined subnetworks. In order to make the subnetworks uniform across the age groups, we repeat the RC and NDI analyses using the population averaged connectome $W_{cohort}$. Furthermore, we apply the $W_{cohort}$ defined subnetworks from both frameworks to each subject's individual connectome for analysis with age. Associations are estimated by fitting a linear model to the subject-level data, given by

$$measure = m * age + b,$$

with slope *m* and offset *b*. Additionally, we report the average strength and density for each subnetwork in both RC and NDI frameworks with corresponding standard deviation.

## Nodal NDI Tier assignment on subject level

Finally, we investigate NDI Tier assignment on a subject level. To do so, we calculate each subject's nodal NDI labelling and then determine the median Tier assignment of each node across 196 connectomes. Furthermore, we quantify the corresponding variation as standard deviation.

All statistical analyses were performed using MATLAB. The code used to calculate the statistics and generate the figures, as well as an implementation for computing NDI is available at https://github.com/mdschirmer/NDI.

# Results

## Network organization by age group

Table 2 details the $k_{max}^{group}$ computed from each group-averaged connectome and the number of corresponding regions determined to form the rich-club subnetwork.

*Table 2: Normalized RC coefficient analysis of the four age groups, identifying the degree range with $\Phi_{group}^{norm}(k)$ significantly greater than 1, $k_{max}^{group}$, and the corresponding number of nodes at $k_{max}^{group}$.*

|  | **U20** | **U40** | **U60** | **O60** |
|---|---|---|---|---|
| **k-range** | 34-49 | 36-48 | 37-47 | 39-53 |
| $k_{max}^{group}$ | 49 | 48 | 47 | 53 |
| **# nodes** | 11 | 10 | 14 | 5 |

In some cases, the atlas assigns multiple nodes to the same anatomical label. Figure 1 shows the rich-club regions identified for each of our four age groups using $k_{max}^{group}$ shown in Table 2. It highlights 14 regions as belonging to the rich-club across the life-span cohort, with eight regions appearing in at least three of the four age groups. These regions consist bilaterally of the caudate, thalamus, pallidum, and parahippocampal posterior regions, as well as the right putamen.

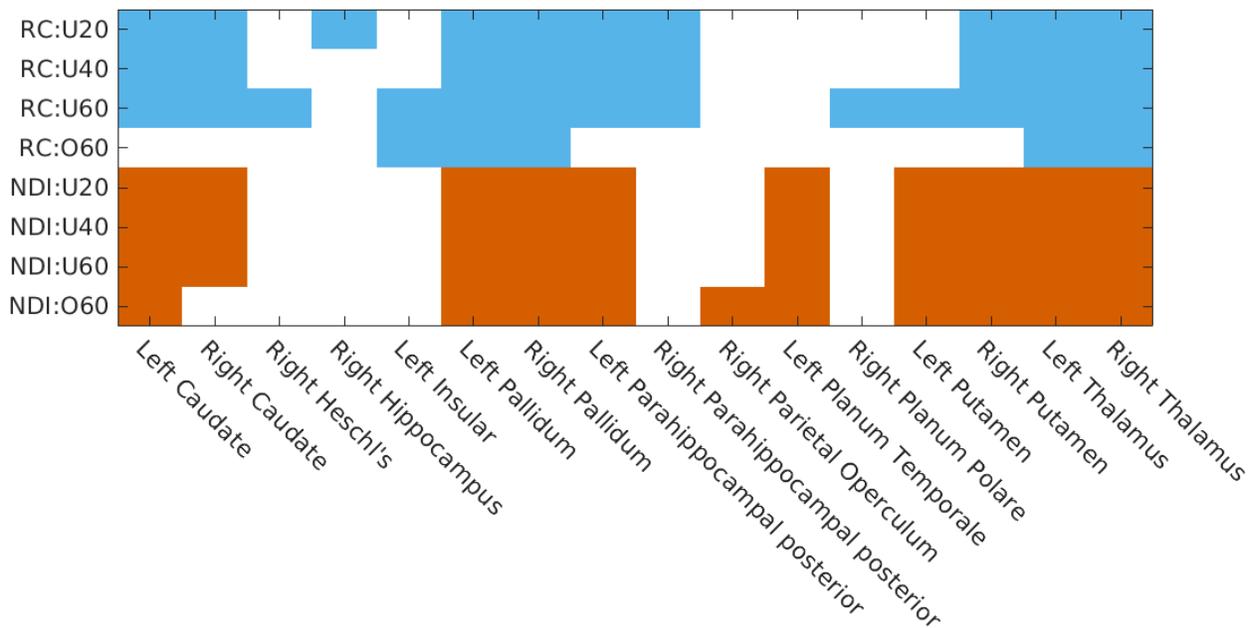

*Figure 1: Rich-club and the top ten brain regions with highest NDI scores defined from the connectome ($W_{group}$) for each age group.*

For NDI, the ten regions with highest score are also indicated in Figure 1. Comparing NDI across the four age groups demonstrates high consistency with an average Spearman's correlation coefficient of 0.98±0.01, with the same correlation in degree sequences between group connectomes.

Using the subnetworks defined by the rich-club, we investigate the NDI scores of RC, F and S. Figure 2 summarizes the NDI scores stratified by RC subnetworks and age groups. RC nodes, on average, have the highest NDI scores of the group connectome, with lower scores for F, followed by S regions.

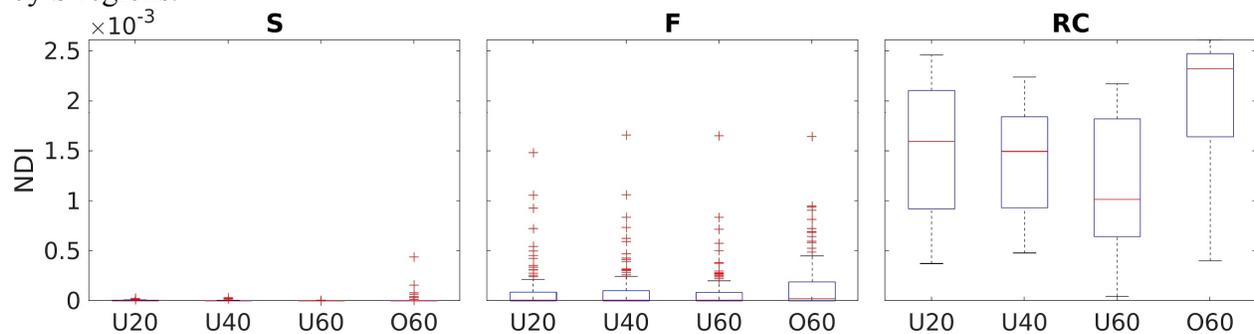

*Figure 2: A comparison of NDI scores stratified by RC subnetworks for each age group.*

## Subnetwork definition using NDI

We observe that each group connectome's NDI follows a mixture of normal distributions after a natural log-transform (Figure 3). AIC and BIC analysis both showed minimas for g = 3, suggesting that three Gaussians described the distribution best (see Figure S1 in Supplementary Materials) and was employed for all further analysis. This resulted in four Tiers in our NDI framework. GMM$_3$ fitted to each of the distributions are shown in Figure 3. The average centers of the Gaussians across all four W$_{group}$ were -14.85 ± 0.67, -11.14 ± 0.08, and -8.19 ± 0.15. The halfway point between consecutive Gaussians (-13.00 and -9.66) divide the NDI distribution into three sections. Subsequently, with the inclusion of the NDI = 0 Tier, each node is assigned to one of the four Tiers according to their NDI value for our analysis.

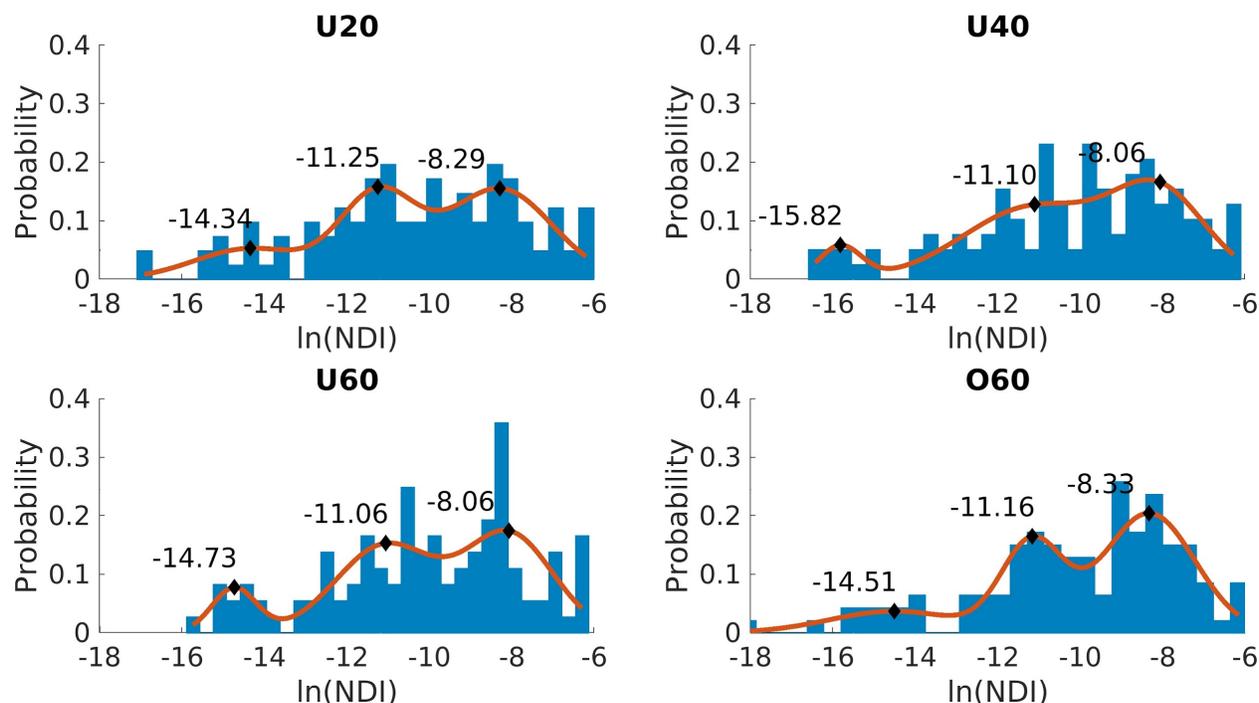

*Figure 3: Natural log-transformed NDI distribution for each age group. Each distribution has a GMM with three Gaussians fitted to it. The centers of these Gaussians are indicated by black diamonds, while the probability density function is shown in orange.*

Figure 4 shows the confusion matrices assessing the stability of nodal assignment for both NDI- and rich-club frameworks across age groups. On average, the largest variations in nodal assignment to a different subnetwork is found with the RC framework (greater off-diagonal percentages in RC confusion matrices, when compared to NDI).

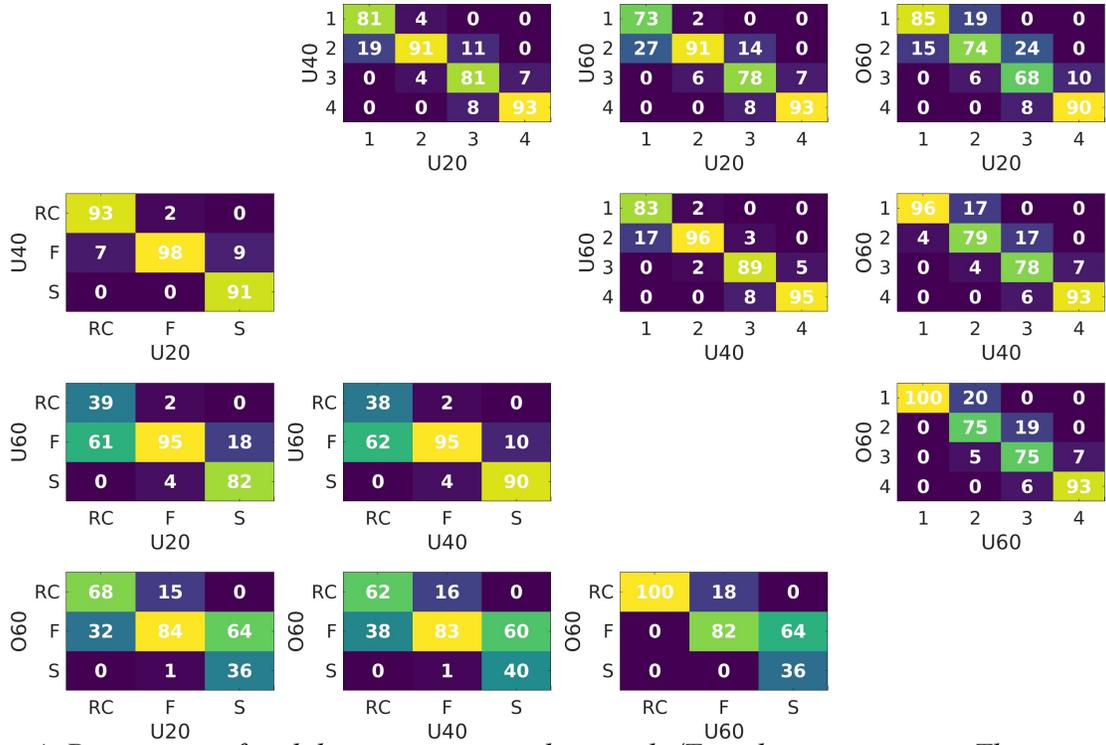

*Figure 4: Percentage of nodal assignment to subnetworks/Tiers between groups. The upper right corresponds to comparisons using the NDI framework, whereas the lower left utilizes the RC framework.*

Network theoretical measures computed from the subnetworks derived via both the RC and NDI frameworks are plotted in Figure 5.

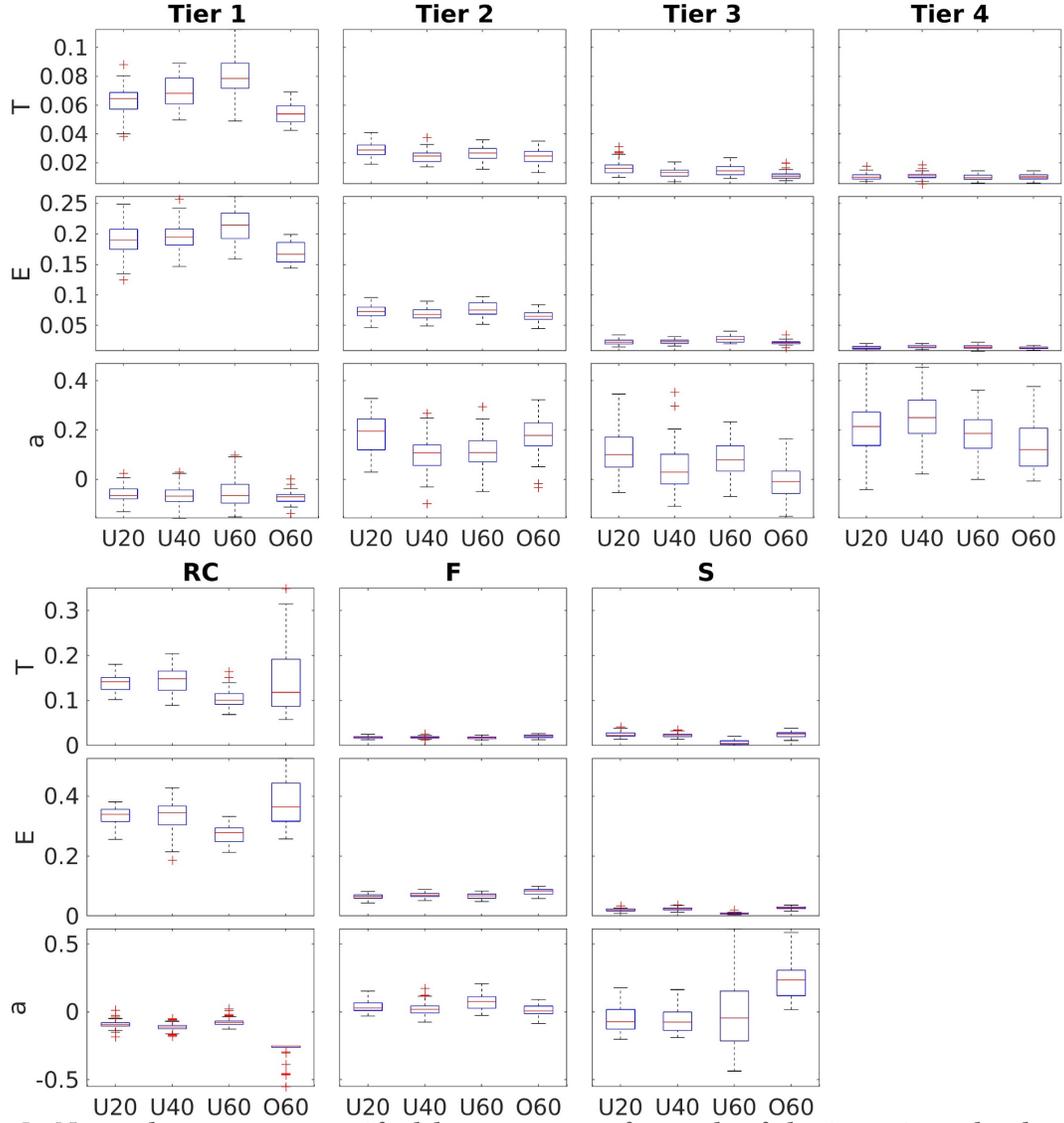

*Figure 5: Network measures stratified by age group for each of the investigated subnetworks. Top: NDI-based subnetworks following their differentiation from fitting a GMM$_3$ model, with decreasing NDI from Tiers 1 to 4. Bottom: RC-based subnetworks, stratified by their connectivity profile with respect to the rich-club.*

## Age-dependent subnetwork associations within the cohort

Figure 6 shows the regions stratified by subnetworks from W$_{cohort}$, as defined from NDI and RC frameworks. NDI nodes were classified according to Gaussian means -14.76,-10.75 and -7.93 for Tiers 1 to 3, respectively, and RC nodes were defined with $k_{max} = 47$.

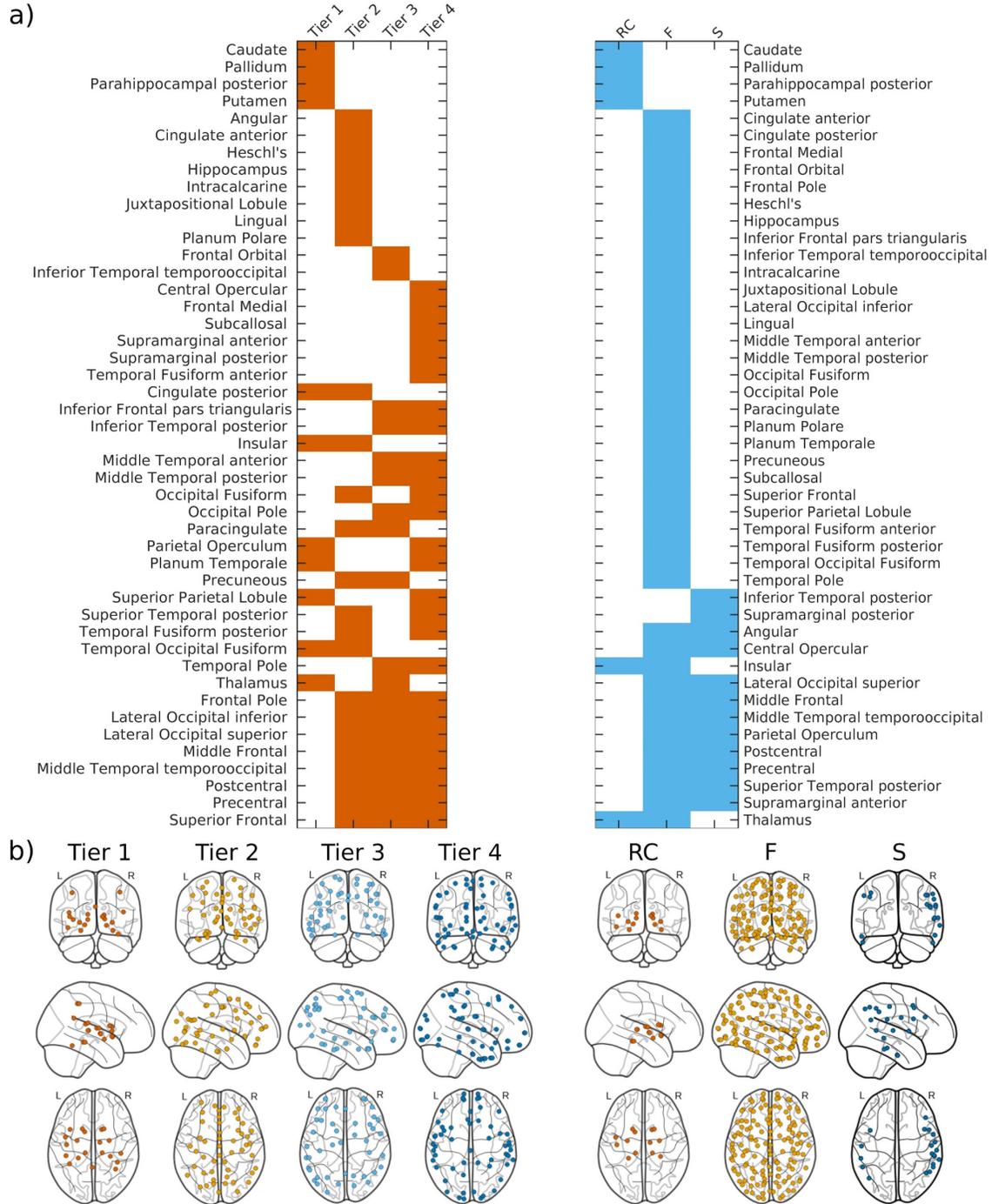

*Figure 6: NDI– (left) and RC-derived (right) subnetworks from cohort-averaged connectome ($W_{cohort}$). a) Assignments of cortical regions by subnetwork for each framework, where corresponding left and right hemispheric, anatomical regions were combined into a single label. Regions are firstly ordered according to their unique membership in a subnetwork, and then by increasing overlap. b) Brain regions plotted by subnetwork label for each framework (enlarged plots are in Figure S2 of Supplementary Materials).*

Figure 7 shows the association of network measures calculated from RC- and NDI-based subnetworks with age. Subnetworks contained 20, 50, 44 and 56 nodes for Tier 1, 2, 3, and 4, with 13, 137, and 20 nodes for RC, F, and S subnetworks respectively. For both frameworks, we see, on average, good separation of each subnetwork investigated in at least one of the three network measures consistently across age. Global efficiency demonstrates the highest separation between subnetworks in both frameworks, with RC and Tier 1 regions having the highest levels of efficiency. We see clear differentiation between regions in the three remaining subnetworks in the NDI-framework (Tiers 2, 3 and 4) , as well as between the two remaining subnetworks in the RC-framework (F and S). This separation between subnetworks can also be observed in terms of transitivity, where Tiers 3 and 4, as well as F and S overlap in part. For assortativity, there is less of an overlap between subnetworks derived from NDI compared to the RC-framework.

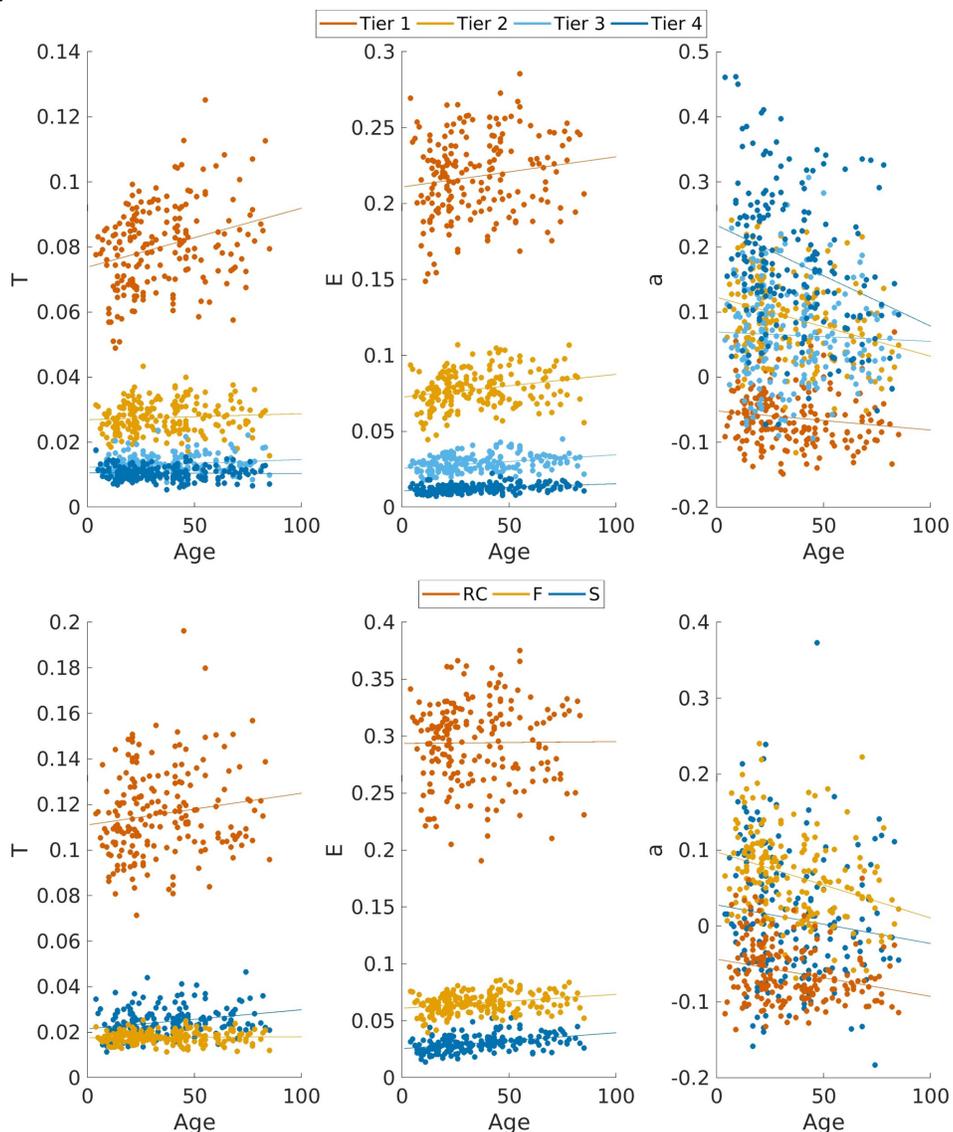

*Figure 7: Network measures transitivity (T), efficiency (E), and assortativity (a) plotted against age for NDI (top) and RC (bottom) subnetworks defined from the full cohort connectome ($W_{cohort}$). Linear regressions lines for each network measure and subnetwork are plotted. NDI Tiers show an average strength of 0.25±0.05, 0.44±0.08, 1.45±0.25, and 2.32±0.33 with an average density of 0.15±0.02, 0.24±0.03, 0.41±0.05, and 0.78±0.07 for Tiers 4, 3, 2, and 1,*

*respectively. RC subnetworks show an average strength of 0.33±0.08, 2.04±0.31, and 2.33±0.33 with an average density of 0.34±0.05, 0.29±0.03, and 0.91±0.07 for RC, F, and S, respectively.*

Fitted parameters of the linear regressions are summarized in Table 3. All subnetworks show increasing efficiency with age, while assortativity decreases. In the RC-based subnetworks, S decreases in transitivity, as does the Tier 4 subnetwork (NDI=0). All other subnetworks exhibit increasing transitivity with age.

*Table 3: Results of linear regression with slope m and offset b for transitivity (T), efficiency (E), and assortativity (a) with age from NDI- (1,2,3,4) and rich-club-based (RC, F, S) subnetworks. Statistically significant regression parameters are in bold (significance at p<0.001, corrected for multiple comparisons).*

|  | T | | | | E | | | | a | | | |
|---|---|---|---|---|---|---|---|---|---|---|---|---|
| **NDI** | 1 | 2 | 3 | 4 | 1 | 2 | 3 | 4 | 1 | 2 | 3 | 4 |
| m | **1.42E-4** | **2.64E-5** | **2.56E-5** | -1.89E-6 | **1.63E-4** | **1.63E-4** | **1.08E-4** | **4.70E-5** | **-3.18E-4** | **-1.00E-3** | **-4.09E-4** | **-1.55E-3** |
| b | 8.07E-2 | 2.88E-2 | 1.29E-2 | 1.06E-2 | 2.19E-1 | 8.08E-2 | 2.82E-2 | 1.09E-2 | -5.42E-2 | 1.40E-1 | 6.75E-2 | 2.34E-1 |
| **RC** | RC | F | S | | RC | F | S | | RC | F | S | |
| m | **1.40E-4** | 4.52E-6 | **-8.39E-5** | | **1.59E-5** | **1.22E-4** | **1.40E-4** | | **-4.90E-4** | **-8.71E-4** | **-5.08E-4** | |
| b | 1.11E-2 | 1.77E-2 | 2.79E-2 | | 2.94E-1 | 6.12E-2 | 2.57E-2 | | -4.35E-2 | 9.77E-2 | 2.79E-2 | |

## Nodal NDI Tier assignment on subject level

Figure 8 shows each node's (region) NDI Tier assignment for every subject, as well as the median assignment with corresponding variation (standard deviation) across all 196 subjects. While some variation exists between Tier assignments on the subject level, both plots indicate relative stability of these assignments on the subject level.

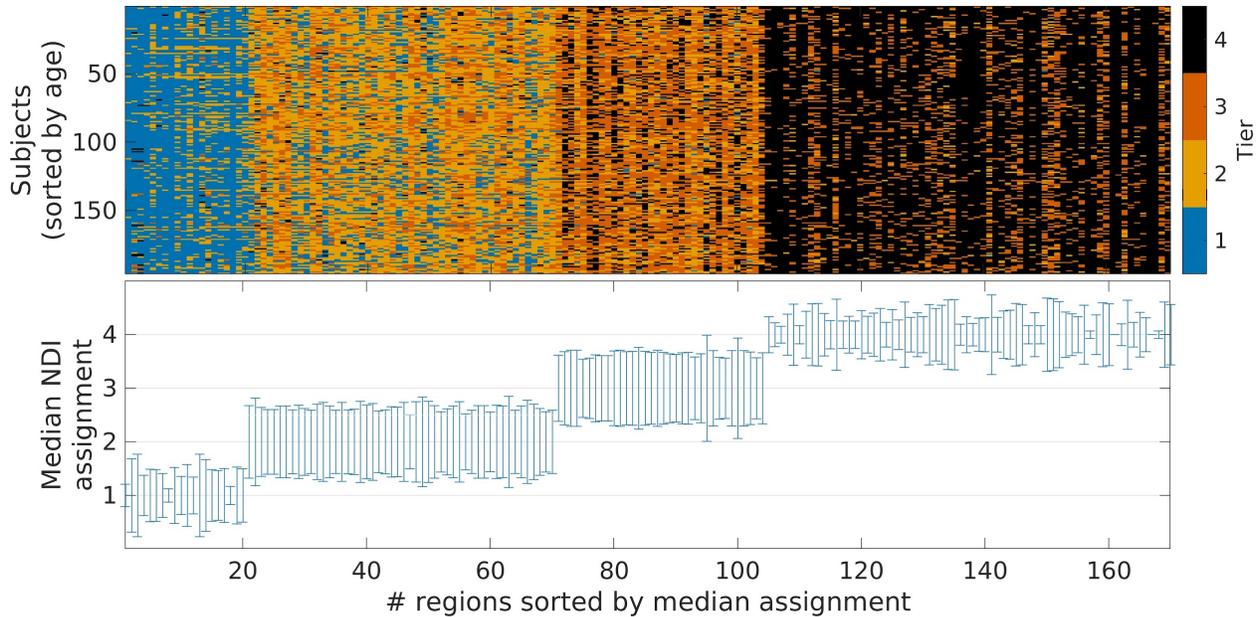

*Figure 8: Top: Assignment to the four NDI Tiers (color coded) for each node and every subject. If an assignment of a given node is stable, the corresponding column will consist of a single color. Subjects are ordered by age from youngest to oldest (top to bottom row). Bottom: Median NDI Tier assignment for each node. Uncertainty is characterized as standard deviation (minimum: 0; maximum: 0.98) of nodal assignment and indicated as error bars.*

# Discussion

In this work we extended the NDI metric into a novel framework to analyse weighted networks and utilized the nodal indices to identify four distinct subnetworks in the human connectome. Importantly, our subnetwork definition is data driven, without any manually chosen parameters for a given connectome, and shows higher consistency across age groups, compared to RC-based subnetworks.

Investigating NDI, stratified by RC-based subnetworks, revealed generally higher NDI for nodes that also belong to the rich-club. This is to be expected, as the rich-club has been shown to be an integral part for effective information transport in networks. Feeders and Seeders have lower NDI, where Seeders almost uniformly exhibit an NDI of zero. This is consistent with the notion that these nodes are 'local' or 'peripheral' in the network. However, we observe that some nodes belonging to Feeder regions exhibit higher importance for network functioning, with NDI values that are comparable to rich-club members. The rich-club is often coined as a backbone for information transport in brain networks. While this aspect has been previously shown, due to the limited assessment of rich-club regions based simply on nodal degree (a localized measure), their importance as a backbone structure is only partially interrogated. NDI, however, investigates the relevance of each region for efficient information transport within the entire brain network by incorporating topological distance information, a more globally informed measure than nodal degree. In doing so, nodal NDI may be more reflective of pathologies or structural changes that implicate entire pathways, making it advantageous over simply counting the number of edges immediately appended to a node.

We observed that NDI followed a mixture of normal distributions after a natural log-transform. Here, we modeled the natural log-transformed NDI of each group connectome and the cohort connectome using a Gaussian Mixture Model into three distributions. By using the halfway point between Gaussian centers (where the probability of a node belonging to either one of the distributions is equal) we distinguish four different subnetworks. Importantly, the Gaussian center estimations are consistent, whether computed by age groups or from the cohort connectome. Furthermore, NDI demonstrated greater stability in identifying more recurring regions with high NDI values in all four age groups than the RC framework (Figure 1). However, the definition of rich-club nodes, due to the lack of consensus on how to define it, is variable. Nonetheless, NDI demonstrates greater stability, even if, e.g., the number of rich-club nodes is fixed across age groups (see Supplementary Material, Figure S3). Stability in identifying regions integral to brain architecture and functioning is desirable for comparative purposes particularly given that they have been similarly detected across species, age, and disease (Ball et al., 2014; Daianu et al., 2015; Grayson et al., 2014; Heuvel and Sporns, 2011; Schirmer and Chung, 2018; van den Heuvel et al., 2013). This has led to a recent uptake in employing *a priori* RC nodes for network analysis (Collin et al., 2016, 2014; van den Heuvel et al., 2013; Wierenga et al., 2018). Utilizing these NDI subnetwork definitions and investigating their network topology with commonly used network measures, we observe distinct patterns for each of the four subnetworks.

Our first analysis investigates age-associated changes in network measures by stratifying the cohort according to age. In both frameworks we observe clear, mostly non-linear patterns in each subnetwork. Similar patterns have been observed previously, e.g. with strength, following an inverse U-shape for RC members (Zhao et al., 2015). However, the membership of each region to a subnetwork (at the RC and Tier 1 level, for example) is not constant with age group in this analysis. While refinement and reassignment of region membership to any subnetwork is possible, comparisons between age groups becomes difficult. In our second experiment, we homogenized the regions that define the subnetworks in both frameworks by utilizing the cohort connectome. Analyzing the trends of the investigated network measures also showed significant patterns associated with age in each subnetwork for both framework. Specifically, we observe a general increase in efficiency with age, most likely reflecting a refinement of the information transport within each subnetwork. It should be noted, however, that whole connectome analyses have identified similar U-shape patterns as mentioned above, warranting further investigation into the differences in network measures computed from subnetworks versus the whole connectome. Assortativity shows a general decrease with age and for all subnetworks, indicating that the connections between nodes of different degrees are strengthened. Transitivity demonstrates mixed patterns for individual subnetworks. It generally increases in RC and F subnetworks, as well as in Tiers 1, 2 and 3, indicative of a within-subnetwork strengthening over age. However, for S and Tier 4, transitivity decreases, reflective of less tightly integrated subnetworks. This may be the result of a distribution of the limited resources in the brain, which favors pre-existing, highly integrated regions. These results suggest that our NDI framework defines organizations of age-associated network principles of T, E and *a* with good separation between subnetworks (Figure 7), indicative of more highly modularized groupings of nodes, when compared with the RC framework, across the ages.

Regarding the functional significance of the NDI-derived subnetworks, it can be clearly noted that the gray matter regions (both subcortical nuclei and cortical regions) follow a unique, largely non-redundant pattern (Figure 6). This is particularly noticeable in separation of the Tier 1 regions from the Tiers 2-4, given that the major components of the Tier 1 subnetwork represent the key relay nuclei in the brain (such as thalamus or the basal ganglia represented by caudate, pallidum, putamen) or cortical regions (such as insular, posterior cingulate gyrus) responsible for processing and redistribution of essential information flow from the primary (motor, sensory, language) cortex to the related association cortices and beyond. Greater overlap appears to exist between the cortical regions represented in Tiers 2-4, which potentially reflects the redundancy necessary for operations involving larger and more complex clusters of gray matter structures. This redundancy may serve to ensure the dynamic connectivity that engages multiple brain functions necessary to sustain defined tasks. Future studies on the functional topography of NDI-derived subnetworks should aim to elucidate the significance and versatility of the tier-specific region interactions, as well as the potential influences that may disrupt or enhance them.

There are some limitations to our study. For our framework, we employed three Gaussians to model the NDI distributions based on our Akaike and Bayesian Information Criteria analysis. While we observe good separation and consistent Gaussian centers across the cohort and four age group connectomes, other choices may be valid depending on the dataset. Additionally, other studies have employed alternative means to investigate versions of the rich-club, such as the participation coefficient, community and/or the distribution index (Grayson et al., 2014; Van Den Heuvel and Sporns, 2011). While these network measures are closely related to NDI, the purpose of our work was not to use NDI as a direct measure of network topology, but to differentiate and identify meaningful subnetworks in the connectome. However, NDI can be utilized as a nodal measure to inform the impact of localized damage, e.g. in conditions such as stroke or brain tumors, and to study differences in network topology, which will be the aim of future studies. Limitations in any life-span analysis include defining the age groups in the cohort, and also how to standardize network estimates across subjects and/or age groups. We utilized age groups to investigate potential changes in subnetwork definition based on differences in the group connectome in the first part of the presented analysis. In particular, subjects in the youngest age group (<20 years; U20) include biological changes such as myelination which can modify the structural connectome globally, as part of normal development. The level in which to subdivide by age during periods of neurological development (or aging) is not trivial. Therefore, in this work we did not focus on these developmental stages, warranting dedicated, fine-grained studies in the U20 group. In terms of standardizing networks for rich-club analysis in a life-span study, identifying consistent RC nodes is complicated due to the lack of consensus on how to choose the degree parameter k. In this analysis, we allowed the RC framework the flexibility to define subnetworks as reflected by the data, by defining RC nodes as $k_{max}^{group}$ for each age group, or propagating the RC nodes from a cohort-averaged network to all subjects. However, NDI consistently demonstrated greater stability in identifying coherent nodes when compared to these RC frameworks. This also holds true if other choices for identifying rich-club nodes are used, e.g., by fixing the number of rich-club nodes across age groups (see Supplementary Material, Figure S3).

In summary, we have demonstrated our NDI framework to successfully identify highly consistent subnetworks across the life-span which can be used to uncover topological aspects of the human connectome. By investigating topology according to subnetworks, we found distinct patterns of increasing efficiency with age and differential changes in transitivity in relation to nodal importance. The combined stability in detecting distinct subnetworks defined using cohort, group through to subject-level connectomes, shows NDI to have great promise for assessing the global implications of a disease in a network across pathologies and developmental time-frames.

# Funding

This project has received funding from the European Union's Horizon 2020 research and innovation programme under the Marie Sklodowska-Curie grant agreement No 753896 (MDS), and the American Heart Association Postdoctoral Fellowship, 19POST34380005 (AWC).

**Table S1:** Number of unique nodes per anatomical label in the Craddock200 atlas.

| Anatomical Label | Number of Nodes | |
|---|---|---|
| | Left Hemisphere | Right Hemisphere |
| Angular | 1 | 1 |
| Caudate | 1 | 1 |
| Cingulate anterior/posterior | 1/2 | 2/2 |
| Central Opercular | 1 | 1 |
| Frontal Medial | 1 | N/A |
| Frontal Orbital | 2 | 1 |
| Frontal Pole | 10 | 10 |
| Hippocampus | 1 | 1 |
| Heschl's | N/A | 1 |
| Insular | 3 | 2 |
| Intracalcarine | N/A | 1 |
| Inferior Frontal pars triangularis | 1 | 1 |
| Inferior Temporal posterior | 2 | 2 |
| Inferior Temporal temporooccipital | 1 | 1 |
| Juxtapositional Lobule | N/A | 1 |
| Lingual | 2 | 2 |
| Lateral Occipital inferior/superior | 2/6 | 3/6 |
| Middle Frontal | 3 | 4 |
| Middle Temporal anterior/posterior | 1/2 | 1/1 |
| Middle Temporal temporooccipital | 1 | 2 |
| Occipital Fusiform | 2 | 1 |
| Occipital Pole | 4 | 3 |
| Pallidum | 1 | 1 |
| Putamen | 1 | 1 |
| Precuneous | 2 | 3 |
| Paracingulate | 2 | 2 |
| Postcentral | 4 | 4 |
| Precentral | 4 | 3 |
| Parahippocampal posterior | 1 | 1 |
| Parietal Operculum | 1 | 2 |
| Planum Temporale | 2 | N/A |
| Planum Polare | N/A | 1 |
| Subcallosal | 1 | N/A |
| Superior Frontal | 3 | 2 |
| Supramarginal anterior/posterior | 1/1 | 1/1 |
| Superior Parietal Lobule | 2 | 2 |
| Superior Temporal posterior | N/A | 2 |
| Thalamus | 2 | 2 |
| Temporal Fusiform anterior/posterior | 1/1 | 1/2 |
| Temporal Occipital Fusiform | 2 | 1 |
| Temporal Pole | 3 | 3 |
| | | |
| **Total number of nodes** | 84 | 86 |

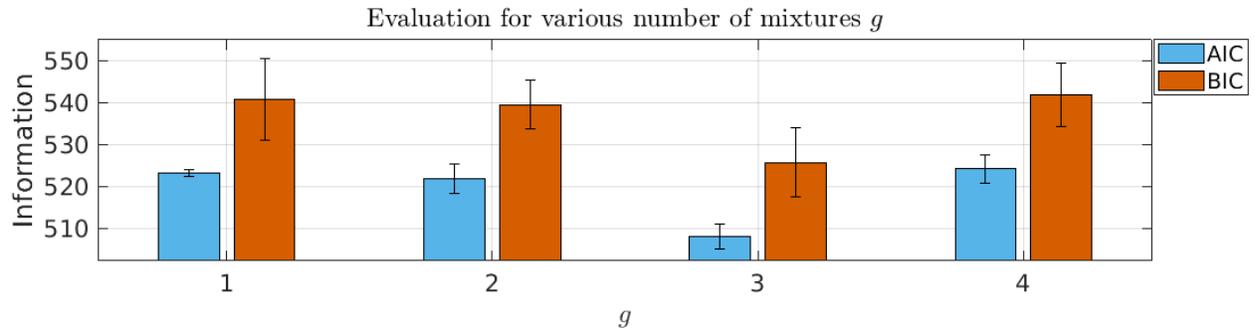

Figure S1: GMM assessment for varying number of Gaussians g fitted to each of the four group connectomes, based on Akaike Information Criterion (AIC) and Bayesian Information Criterion (BIC). Both measure suggest that g=3 best describes the observed data.

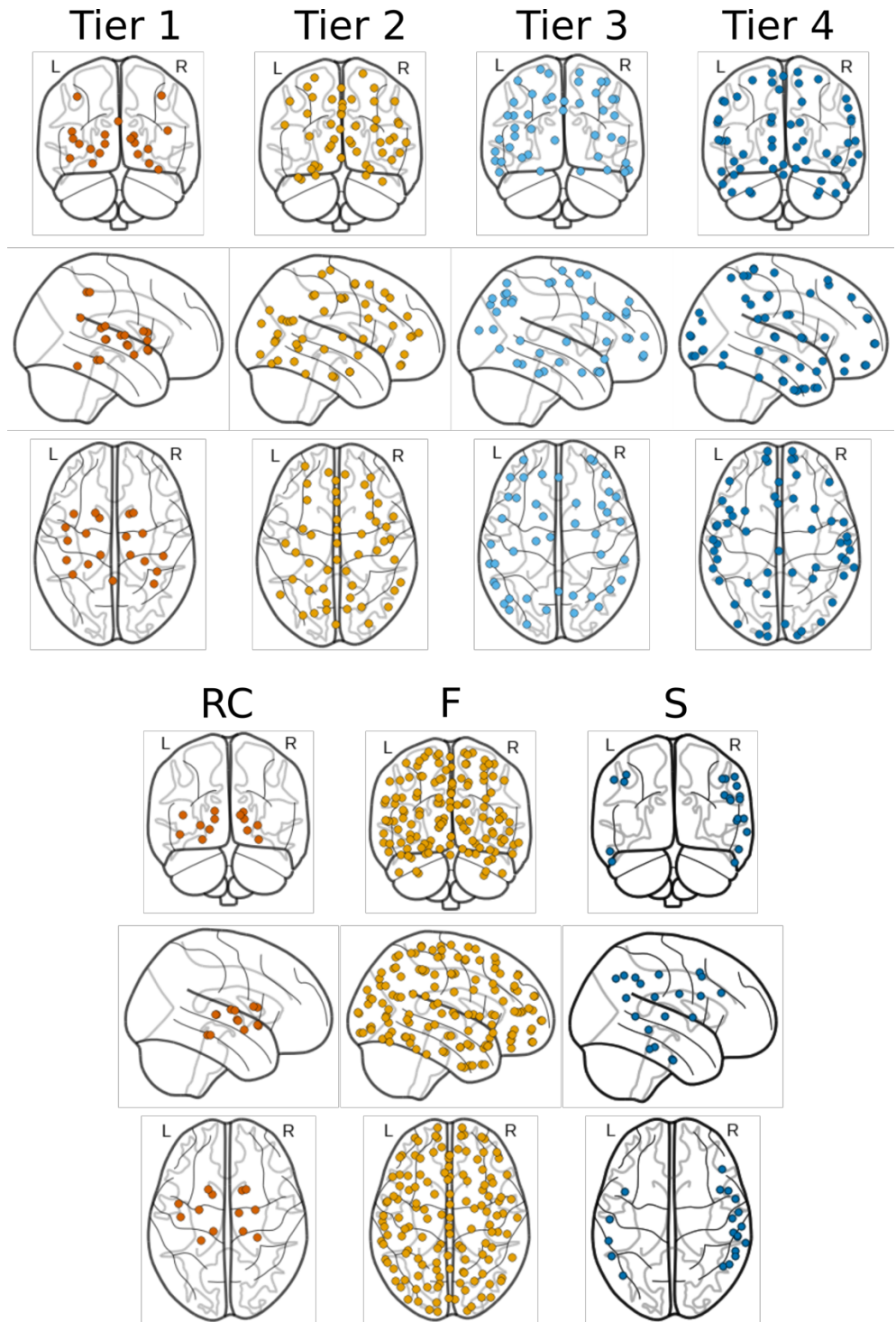

Figure S2: Brain regions plotted by subnetwork label for each framework. Top: NDI framework with four Tiers (decreasing importance). Bottom: RC framework with rich-club (RC), Feeder (F), and Seeder (S) subnetworks.

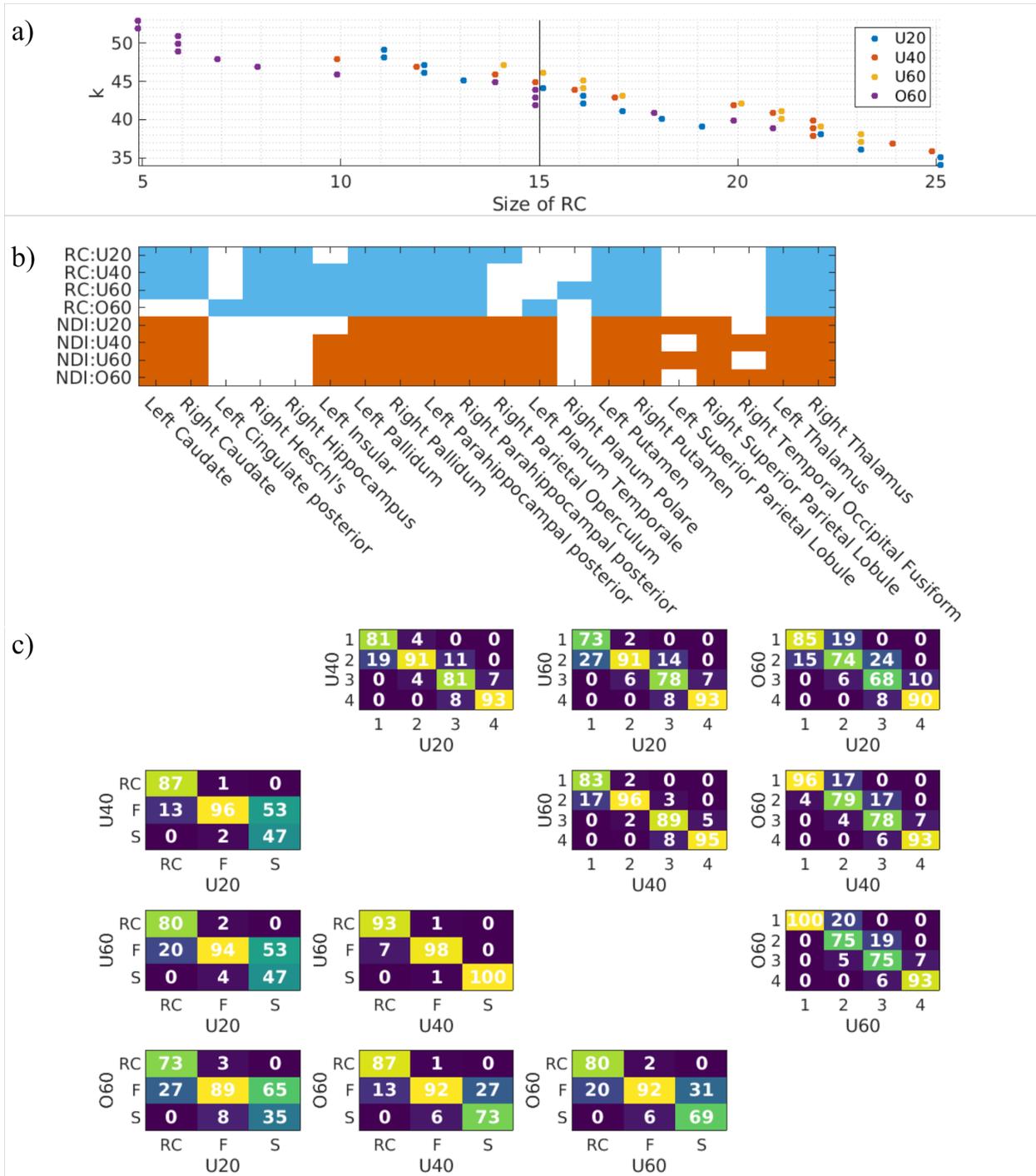

Figure S3: Analysis with consistent number of rich-club nodes between age groups, but variable choice of k (44, 45, 46, and 44 for U20, U40, U60, and O60, respectively). a) Analysis of size of rich-club with varying degree k for each age group. The graph demonstrates that consistent rich-club size analysis is only possible for 15 nodes in this data set. b) Rich-club and the top 15 brain regions with highest NDI scores. c) Percentage of nodal assignment to subnetworks/Tiers between groups. The upper right corresponds to comparisons using the NDI framework, whereas the lower left utilizes the RC framework with 15 nodes.